\documentclass[11pt]{article}
\usepackage{amsmath,amsxtra}
\usepackage{amscd}
\usepackage{graphicx}
\usepackage{latexsym}
\usepackage{amsmath,amsthm,amsfonts,amssymb,cite}

\usepackage{latexsym}
\usepackage{amsmath}

\theoremstyle{remark}

\newtheorem*{rem}{Remark}
\numberwithin{equation}{section}
\begin{document}
\hoffset = -2.4truecm \voffset = -2truecm
\renewcommand{\baselinestretch}{1.2}
\newcommand{\mb}{\makebox[10cm]{}\\ }

\newtheorem{theorem}{Theorem}
\newtheorem{proposition}{Proposition}
\newtheorem{lemma}{Lemma}
\newtheorem{definition}{Definition}


\title{Integrable generalizations of the two new soliton hierarchies of AKNS and KN types associated with so(3,$\mathbb{R}$)}
\author{Chun-Xia Li$^{1}$\footnote{trisha\_li2001@163.com}, Shou-Feng Shen$^{2}$, Wen-Xiu Ma$^{3}$ and Shui-Meng Yu$^{4}$\\
$^{1}$School of Mathematical Sciences,
Capital Normal University, \\
Beijing 100048, PR China\\
$^{2}$Department of Applied Mathematics, Zhejiang University of Technology,\\
Hangzhou 310023, PR China\\
$^{3}$Department of Mathematics and Statistics, University of South Florida,\\
Tampa, FL 33620-5700, USA\\
$^{4}$ School of Sciences, Jiangnan University, Wuxi 214122, PR China
}
\date{}
 \maketitle
\begin{abstract}
 The two matrix spectral problems of Ablowitz-Kaup-Newell-Segur (AKNS) and Kaup-Newell (KN) types associated with so(3,$\mathbb{R}$) are generalized. The corresponding hierarchies of generalized soliton equations
  are derived by the standard procedure using the zero curvature formulation.
   Recursion operators and bi-Hamiltonian structures
   are explicitly constructed for the resulting two generalized soliton hierarchies of AKNS and KN types, which shows their Liouville integrability.
\end{abstract}

\indent\indent\indent\indent\indent\indent  bi-Hamiltonian structures, Liouville integrability

\section{Introduction}
One important and interesting topic in soliton theory is to
search for new integrable systems and investigate their integrability properties.
There has been a lot of work on how to generate soliton hierarchies from matrix spectral problems or Lax pairs by the standard procedure
using the zero curvature formulation
\cite{AC,NM,CH,DS,ASF}. The celebrated examples include the Korteweg-de Vries hierarchy, the Ablowitz-Kaup-Newell-Segur (AKNS) hierarchy, the
Kaup-Newell (KN) hierarchy, the Wadati-Konno-Ichikawa (WKI) hierarchy, the Dirac hierarchy and the Boiti-Pempinelli-Tu hierarchy \cite{PDL,AKNS,KN,WKI,DR,BPT}. Soliton hierarchies generated from matrix spectral problems often possess recursion operators and
bi-Hamiltonian structures, which implies Liouville integrability.
The associated bi-Hamiltonian structures can be established by the trace identity or the variational identity \cite{TGZ,MA4}.
Which identity to use depends
 on whether the underlying matrix loop algebra is semisimple or not. If there exist bi-Hamiltonian structures, the associated
  Hamiltonian pairs often generate hereditary recursion operators \cite{FMR,CDO,FF,PO}.

Recently, one of the authors (Ma) successfully generalized the classical AKNS and KN
matrix spectral problems associated with the special linear algebra sl(2,$\mathbb{R}$) to the ones associated with the special orthogonal
 Lie algebra so(3,$\mathbb{R}$), derived the corresponding hierarchies of commuting soliton equations and established
 their bi-Hamiltonian structures \cite{MA1,MA2},
 by the standard procedure using the zero curvature formulation \cite{TGZ,MA6}.

The three-dimensional real special orthogonal Lie algebra so(3,$\mathbb{R}$) consists of $3\times 3$ skew-symmetric matrices. This Lie algebra is simple and has the basis
\begin{equation}
e_1=\begin{pmatrix}0&0&-1\\ 0&0&0\\ 1&0&0\end{pmatrix},\ e_2=\begin{pmatrix}0&0&0\\ 0&0&-1\\ 0&1&0\end{pmatrix},\ e_3=\begin{pmatrix}0&-1&0\\ 1&0&0\\ 0&0&0\end{pmatrix},
\end{equation}
with the commutator relations:
\begin{align*}
[e_1,e_2]=e_3,\  [e_2,e_3]=e_1,\  [e_3,e_1]=e_2.
\end{align*}
The derived algebra of so(3,$\mathbb{R}$) is so(3,$\mathbb{R}$) itself, and so(3,$\mathbb{R}$) is one of the only two three-dimensional real Lie algebras with a three-dimensional derived algebra. The other one is the special linear algebra sl(2,$\mathbb{R}$),
 which has been widely used in soliton theory. The Lie algebra sl(2,$\mathbb{R}$) has the basis
\begin{equation}
e_1'=\begin{pmatrix}1&0\\ 0&-1\end{pmatrix},\ e_2'=\begin{pmatrix}0&1\\ 0&0\end{pmatrix},\ e_3'=\begin{pmatrix}0&0\\ 1&0\end{pmatrix},
\end{equation}
which has the following commutator relations:
\begin{align*}
[e_1',e_2']=2e_2',\ [e_3',e_1']=2e_3',\ [e_2',e_3']=e_1'.
\end{align*}

Let $u=(p,q)^T$ and denote the spectral parameter by $\lambda$.
 The classical AKNS matrix spectral problem reads
\begin{align}
\phi_x=(-\lambda e_1'+p e_2'+qe_3')\phi.
\end{align}
and the classical KN matrix spectral problem,
\begin{align}
\phi_x=(\lambda^2 e_1'+\lambda pe_2'+\lambda qe_3')\phi.
\end{align}
In \cite{MA1,MA2}, Ma generalized the above matrix spectral problems to the ones associated with so(3,$\mathbb{R}$):
\begin{align}\label{AKN1}
\phi_x=(-\lambda e_1+pe_2-qe_3)\phi,
\end{align}
and
\begin{align}\label{KN1}
\phi_x=(\lambda^2 e_1+\lambda pe_2+\lambda qe_3)\phi,
\end{align}
taking the same linear combination of the basis vectors.
Stimulated by the generalized AKNS and KN matrix spectral problems in \cite{GengMa,YZY},
we would like to generalize the two new matrix spectral problems \eqref{AKN1} and \eqref{KN1} associated with so(3,$\mathbb{R}$) similarly.
Our two generalizations of \eqref{AKN1} and \eqref{KN1} are
\begin{align}\label{AKNSS}
\phi_x=[-(\lambda+\alpha (p^2+q^2) )e_1+pe_2-qe_3]\phi,
\end{align}
and
\begin{align}\label{KNS}
\phi_x=[(\lambda^2+\alpha(p^2+q^2))e_1+\lambda pe_2+\lambda qe_3)]\phi,
\end{align}
where $\alpha$ is an arbitrary constant.
Obviously, the generalized matrix spectral problems \eqref{AKNSS} and \eqref{KNS} with the case of $\alpha=0$
reduce to \eqref{AKN1} and \eqref{KN1}, respectively.
In this sense, we call \eqref{AKNSS} a generalized Ma matrix spectral problem of AKNS type
associated with so(3,$\mathbb{R}$) and \eqref{KNS} a generalized Ma matrix spectral problem of KN type
associated with so(3,$\mathbb{R}$).

The paper is structured as follows. In Section 2 and Section 3,
 from the two generalized matrix spectral problems \eqref{AKNSS} and \eqref{KNS},
 we shall derive two soliton hierarchies and construct their recursion operators and bi-Hamiltonian structures, thereby proving their Liouville integrability. In Section 4, a conclusion and discussion will be given.

\section{Generalized Ma equations of AKNS type associated with so(3,$\mathbb{R}$)}

\subsection{A hierarchy of generalized Ma equations of AKNS type}
To generate a hierarchy of generalized Ma equations of AKNS type from \eqref{AKNSS},
let us introduce the spectral matrix
\begin{align}
U=U(u,\lambda)=
\begin{pmatrix}0&q&\lambda+\alpha(p^2+q^2)\\-q&0&-p\\-\lambda-\alpha(p^2+q^2)&p&0\end{pmatrix}
,\ u
=\begin{pmatrix}p\\q\end{pmatrix},
\end{align}
where $\lambda $ is the spectral parameter, and so,
the generalized matrix spectral problem \eqref{AKNSS} becomes
\begin{align}
\phi_x=[-(\lambda+\alpha(p^2+q^2))e_1-pe_2+qe_3)]\phi=U\phi=
U(u,\lambda )
\phi,\ \phi=\begin{pmatrix}\phi_1\\\phi_2\\\phi_3\end{pmatrix}.
\end{align}


Following the standard procedure using the zero curvature formulation \cite{TGZ,MA6},
 let us first solve the stationary zero curvature equation
\begin{equation}
W_x=[U,W],\ W=\begin{pmatrix}0&c&a\\-c&0&-b\\-a&b&0\end{pmatrix}
\end{equation}
which gives
\begin{equation}\left\{\begin{array}{l}\label{SZCE1}
a_x=pc-qb,\\
b_x=qa-\lambda c-\alpha(p^2+q^2)c,\\
c_x=\lambda b-pa+\alpha(p^2+q^2)b.\end{array}\right.
\end{equation}
Upon letting \begin{equation}\label{CEF}a=\sum\limits_{i\ge 0}a_i\lambda^{-i},\ b=\sum\limits_{i\ge 0}b_i\lambda^{-i},\ c=\sum\limits_{i\ge 0}c_i\lambda^{-i},\end{equation} and taking initial values
\begin{align}
a_0=-1,\ b_0=c_0=0,
\end{align}
the systems \eqref{SZCE1} leads to
\begin{equation}
\left\{
\begin{array}{l}
a_{ix}=pc_i-qb_i,\\
b_{i+1}=c_{i,x}+pa_i-\alpha(p^2+q^2)b_i,\\
c_{i+1}=-b_{i,x}+qa_i-\alpha(p^2+q^2)c_i,
\end{array}
\right.\ i\ge 0,
\end{equation}
which tells the recursion relations:
\begin{align}
\begin{pmatrix}
b_{i+1}\\
c_{i+1}
\end{pmatrix}
&=L\begin{pmatrix}
b_{i}\\c_{i}
\end{pmatrix}=\begin{pmatrix}-\alpha(p^2+q^2)-p\partial^{-1}q&\partial+p\partial^{-1}p\\
-\partial-q\partial^{-1}q&q\partial^{-1}p-\alpha(p^2+q^2)
\end{pmatrix}
\begin{pmatrix}
b_{i}\\c_{i}
\end{pmatrix},\ i\ge 0.
\end{align}
By imposing the conditions on the constants of integration
\begin{align}
a_i|_{u=0}=b_i|_{u=0}=c_i|_{u=0}=0,\,\ i\ge 1,\label{condsofConstantsofIntegration}
\end{align}
we can determine the sequence $\{a_i,b_i,c_i|\,i\ge 1\}$ uniquely. In this way, we can computer the first few sets as follows:
\begin{align*}
&a_1=0,\ b_1=-p,\ c_1=-q;\\
&a_2=\frac{1}{2}(p^2+q^2),\ b_2=-q_x+\alpha p(p^2+q^2),\
c_2=p_x+\alpha q(p^2+q^2);\\
&a_3=pq_x-qp_x-\alpha(p^2+q^2)^2,\\
&b_3=p_{xx}+2\alpha(p^2+2q^2)q_x+2\alpha pqp_x-\alpha^2 p(p^2+q^2)^2+\frac{1}{2}p(p^2+q^2),\\
&c_3=q_{xx}-2\alpha(q^2+2p^2)p_x-2\alpha pqq_x-\alpha^2q(p^2+q^2)^2+\frac{1}{2}q(p^2+q^2).
\end{align*}

Let us now introduce the auxiliary spectral problems
\begin{align}\label{AKNSA}
\phi_{t_m}&=V^{[m]}\phi,\,V^{[m]}=(\lambda^mW)_++\Delta_m=(\lambda^mW)_++\begin{pmatrix}0&0&f_m\\
0&0&0\\
-f_m&0&0\end{pmatrix},\  m\ge 0,
\end{align}
where $P_+$ denotes the polynomial part of $P$ in $\lambda$ and $f_m,\ m\ge 0$, satisfy
\begin{align}\label{FE1}
f_{m,x}=2\alpha(pp_{t_m}+qq_{t_m}),\ m\ge 0.
\end{align}
The compatibility conditions of \eqref{AKNSS} and \eqref{AKNSA} with each $m\ge 0$, i.e.,
 the zero curvature equations
\begin{align}
U_{t_m}-V^{[m]}_x+[U,V^{[m]}]=0,\ m\ge 0,
\end{align}
 yield
\begin{equation}\left\{\begin{array}{l}
p_{t_m}=b_{m,x}-qf_m-qa_m+\alpha(p^2+q^2)c_m=-c_{m+1}-qf_m,\\
q_{t_m}=c_{m,x}+pf_m+pa_m-\alpha(p^2+q^2)b_m=b_{m+1}+pf_m,\end{array}\right.\ m\ge 0.\label{ResultingEquations}
\end{equation}
Based on these systems \eqref{ResultingEquations}, we can solve \eqref{FE1} to get
\begin{align}
f_m=2\alpha\partial^{-1}(qb_{m+1}-pc_{m+1})=-2\alpha a_{m+1},\ m\ge 0.\label{defoff_m}
\end{align}
Therefore, upon plugging
\eqref{defoff_m} into \eqref{ResultingEquations},
we finally arrive at
the hierarchy of generalized Ma equations of AKNS type:
\begin{align}\label{SH1}
u_{t_m}
=K_{m}=R\begin{pmatrix}
b_{m+1}\\c_{m+1}
\end{pmatrix}, \ R=\begin{pmatrix}
-2\alpha q\partial^{-1}q&-1+2\alpha q\partial^{-1}p\\
1+2\alpha p\partial^{-1}q&-2\alpha p\partial^{-1}p
\end{pmatrix},
\ m\ge 0.
\end{align}
The first nonlinear system in the soliton hierarchy \eqref{SH1} is as follows:
\begin{align}
u_{t_2}=\begin{pmatrix}p\\q\end{pmatrix}_{t_2}=\begin{pmatrix}-q_{xx}+4\alpha pqq_x+4\alpha p^2p_x-\alpha^2 q(p^2+q^2)^2-\frac{1}{2}q(p^2+q^2)\\
p_{xx}+4\alpha pqp_x+4\alpha q^2q_x+\alpha^2 p(p^2+q^2)^2+\frac{1}{2}p(p^2+q^2)\end{pmatrix}.
\end{align}

\subsection{Bi-Hamiltonian structures and Liouville integrability of the first generalized soliton hierarchy}

In order to establish bi-Hamiltonian structures, we shall use the trace identity \cite{TGZ} [or more generally the variational identity (see, e.g., \cite{MA5,MaMZ-GJMS2012} for details)]:
\begin{equation}\label{TIF}
\frac{\delta}{\delta u}\int\mbox{tr}\left(W\frac{\partial U}{\partial \lambda}\right)dx=\left(\lambda^{-\gamma}\frac{\partial}{\partial \lambda}\lambda^{\gamma}\right)
\left(\mbox{tr}\left(W\frac{\partial U}{\partial u}\right)\right),
\end{equation}
where $\gamma$ is a constant to be determined. It is direct to calculate
\begin{align}\label{TIFF}
\mbox{tr}\left(W\frac{\partial U}{\partial \lambda}\right)=-2a,\ \mbox{tr}\left(W\frac{\partial U}{\partial p}\right)=-2b-4\alpha pa,\ \mbox{tr}\left(W\frac{\partial U}{\partial q}\right)=-2c-4\alpha qa.
\end{align}
By substituting \eqref{CEF} and \eqref{TIFF} into \eqref{TIF} and balancing coefficients of each power of $\lambda$, we have
\begin{align}
\frac{\delta}{\delta u}\left(\int a_{m+1}dx\right)=(\gamma-m)\begin{pmatrix}b_m+2\alpha pa_m\\c_m+2\alpha qa_m\end{pmatrix},\ m\ge 0.
\end{align}
To fix the constant $\gamma$, we can simply let $m=1$ in the above equation and get $\gamma=0$. Thus, we have
\begin{align}\frac{\delta \mathcal{H}_{m}}{\delta u}
=
\begin{pmatrix}b_{m}+2\alpha pa_{m}\\c_{m}+2\alpha qa_{m}\end{pmatrix},\
\mathcal{H}_{m}
=\int \left(-\frac{ a_{m+1}}{m}\right)dx
,\  m\ge 1.
\end{align}


Noticing that
\begin{align}
\begin{pmatrix}b_{m+1}\\c_{m+1}\end{pmatrix}=N\begin{pmatrix}b_{m+1}+2\alpha pa_{m+1}\\c_{m+1}+2\alpha qa_{m+1}\end{pmatrix},\ 
N=\begin{pmatrix}1+2\alpha p\partial^{-1}q&-2\alpha p\partial^{-1}p\\2\alpha q\partial^{-1}q&1-2\alpha q\partial^{-1}p\end{pmatrix}
,\ m\ge 0,
\end{align}
we find that
\begin{align}
u_{t_m}&=K_{m}=J\begin{pmatrix}b_{m+1}+2\alpha pa_{m+1}\\c_{m+1}+2\alpha qa_{m+1}\end{pmatrix}
\end{align}
where
\begin{equation}
J=RN
=\begin{pmatrix}
-4\alpha q\partial^{-1}q&-1+4\alpha q\partial^{-1}p\\
1+4\alpha p\partial^{-1}q&-4\alpha p\partial^{-1}p
\end{pmatrix},
\end{equation}
which is a Hamiltonian operator. It follows now that the soliton hierarchy \eqref{SH1} has the Hamiltonian structures:
\begin{equation}
u_{t_m}=K_m=J\frac{\delta \mathcal{H}_{m+1}}{\delta u},\,m\ge 0
\end{equation}
with
\begin{align}
\mathcal{H}_0=-\alpha\int(p^2+q^2)dx,\, \mathcal{H}_{m}=\int\left(-\frac{a_{m+1}}{m}\right)dx,\ m\ge 1.
\end{align}

It is obvious that
\begin{align}
\frac{\delta \mathcal{H}_{m+1}}{\delta u}=\Psi\frac{\delta \mathcal{H}_m}{\delta u},\ \Psi=N^{-1}LN,
\end{align}
where the inverse operator of $N^{-1}$ is given by
\begin{equation}
N^{-1}=\begin{pmatrix}1-2\alpha p\partial^{-1}q&2\alpha p\partial^{-1}p\\-2\alpha q\partial^{-1}q&1+2\alpha q\partial^{-1}p\end{pmatrix}.
\end{equation}
Then from $K_{m+1}=\Phi K_m,\,\  m\ge 0$, and $J\Psi=\Phi J$, we obtain a common recursion operator for the generalized
soliton hierarchy \eqref{SH1}:
\begin{equation}\Phi=\Psi^{\dagger}=N^{\dagger}L^{\dagger}(N^{-1})^{\dagger}, \label{Phiof1stgsh}\end{equation}
where $\Psi^{\dagger}$ denotes the adjoint operator of $\Psi$.
%
The operator $\Phi=(\Phi_{ij})_{2\times 2}$ can be explicitly computed as follows:
\begin{equation}
\left\{\begin{array}{l}
\Phi_{11}=-2\alpha \partial p\partial^{-1}p+2\alpha q\partial^{-1}q(\partial+2\alpha \partial q\partial^{-1}p)\\
\quad\quad\quad-\alpha (p^2+q^2)(1+2\alpha q\partial^{-1}p)+q\partial^{-1}p[1+2\alpha^2(p^2+q^2)+4\alpha^2\partial p\partial^{-1}p],\\
\Phi_{12}=(1-2\alpha q\partial^{-1}p)(\partial-2\alpha \partial p\partial^{-1}q) \\ \quad\quad\quad-2\alpha^2(p^2+q^2)q\partial^{-1}q+q\partial^{-1}q[1+2\alpha^2(p^2+q^2)+4\alpha^2\partial q\partial^{-1}q],\\
\Phi_{21}=-(1+2\alpha p\partial^{-1}q)(\partial+2\alpha \partial q\partial^{-1}p)\\
\quad\quad\quad+2\alpha^2(p^2+q^2)p\partial^{-1}p-p\partial^{-1}p[1+2\alpha^2(p^2+q^2)+4\alpha^2\partial p\partial^{-1}p],\\
\Phi_{22}=-2\alpha \partial q\partial^{-1}q+2\alpha p\partial^{-1}p(\partial-2\alpha \partial p\partial^{-1}q)\\
\quad\quad\quad-\alpha(p^2+q^2)(1-2\alpha p\partial^{-1}q)-p\partial^{-1}q[1+2\alpha^2(p^2+q^2)+4\alpha^2\partial q\partial^{-1}q].\end{array}\right.
\end{equation}

 It is now a direct computation that all members in the generalized soliton hierarchy \eqref{SH1} are bi-Hamiltonian: 
\begin{equation}
u_{t_m}=K_m=J\frac{\delta \mathcal{H}_{m+1}}{\delta u}=M\frac{\delta \mathcal{H}_{m}}{\delta u},\ m\ge 0,\label{biHSof1stgsh}
\end{equation}
where the second Hamiltonian operator $M$ is given by
\begin{align}
M=\Phi J=\begin{pmatrix}M_{11}&M_{12}\\ M_{21}&M_{22}\end{pmatrix}
\end{align}
with the entries of $M$ being defined by
\begin{equation}\left\{\begin{array}{l}
M_{11}=\partial+2\alpha\partial p\partial^{-1}q+2\alpha^2(p^2+q^2)q\partial^{-1}q\\
\qquad\quad-2\alpha q\partial^{-1}p(\partial-2\alpha \partial p\partial^{-1}q)+q\partial^{-1}q[1+2\alpha^2(p^2+q^2)-4\alpha^2\partial q\partial^{-1}q],\\
M_{12}=-2\alpha \partial p\partial^{-1}p+\alpha(p^2+q^2)(1-2\alpha q\partial^{-1}p)\\
\qquad\quad+2\alpha q\partial^{-1}q(2\alpha\partial q\partial^{-1}p-\partial)-q\partial^{-1}p[1+2\alpha^2(p^2+q^2)-4\alpha^2\partial p\partial^{-1}p],\\
M_{21}=2\alpha \partial q\partial^{-1}q-\alpha(p^2+q^2)(1+2\alpha p\partial^{-1}q)\\
\qquad\quad+2\alpha p\partial^{-1}p[\partial+2\alpha \partial p\partial^{-1}q]+p\partial^{-1}q[-1-2\alpha^2(p^2+q^2)+4\alpha^2\partial q\partial^{-1}q],\\
M_{22}=\partial-2\alpha\partial q\partial^{-1}p+2\alpha^2(p^2+q^2)p\partial^{-1}p\\
\qquad\quad+2\alpha p\partial^{-1}q(\partial-2\alpha \partial q\partial^{-1}p)+p\partial^{-1}p[1+2\alpha^2(p^2+q^2)-4\alpha^2\partial p\partial^{-1}p].\end{array}\right.
\end{equation}
So far, we have established the bi-Hamiltonian structures and therefore proved the Liouville integrability of the first generalized Ma soliton hierarchy \eqref{SH1}. Further, it follows that there are infinitely many commuting common symmetries and conserved functionals
\begin{align}
[K_l,K_m]=K_l'(u)[K_m]-K_m'(u)[K_l]=0,\ l,m\ge 0
\end{align}
and
\begin{equation}\left\{\begin{array}{l}
\{\mathcal{H}_l,\mathcal{H}_m\}_{J}=\int ( \frac{\delta\mathcal{H}_l}{\delta u})^T J\frac{\delta \mathcal{H}_m}{\delta u}dx=0,\\
\{\mathcal{H}_l,\mathcal{H}_m\}_{M}=\int ( \frac{\delta\mathcal{H}_l}{\delta u})^T R\frac{\delta \mathcal{H}_m}{\delta u}dx=0,\end{array}\right. \ l, m\ge 0.
\end{equation}
\begin{rem}
By taking $\alpha=0$ for the generalized matrix spectral problem \eqref{AKNSS},
  the recursion operator \eqref{Phiof1stgsh} and bi-Hamiltonian structures \eqref{biHSof1stgsh}
  of the corresponding generalized soliton hierarchy \eqref{SH1}
  are reduced to the ones
  of the
  Ma soliton hierarchy of AKNS type
           in \cite{MA1}. 
\end{rem}

\section{Generalized Ma equations of KN type associated with so(3,$\mathbb{R}$)}
\subsection{A hierarchy of generalized Ma equations of KN type}

In the same way as shown in Section 2, we can construct a hierarchy of generalized Ma equations of KN type
from the generalized matrix spectral problem \eqref{KNS}, i.e.,
\begin{align}\label{KNS1}
\phi_x=U\phi=[(\lambda^2+\alpha(p^2+q^2))e_1+\lambda pe_2+\lambda qe_3)]\phi,\ u=\begin{pmatrix}p\\q\end{pmatrix},\ \phi=\begin{pmatrix}\phi_1\\ \phi_2\\ \phi_3\end{pmatrix}
\end{align}
with the spectral matrix $U$ being defined by
\begin{align}
U=U(u,\lambda )=\begin{pmatrix}0&-\lambda q&-\lambda^2-\alpha(p^2+q^2)\\
\lambda q&0&-\lambda p\\
\lambda^2+\alpha(p^2+q^2)&\lambda p&0
\end{pmatrix}.
\end{align}

Let us first begin with solving the stationary zero curvature equation
\begin{align}
W_x=[U,W], \ W=\begin{pmatrix}0&-c&-a\\c&0&-b\\a&b&0\end{pmatrix}
\end{align}
which gives
\begin{equation}\left\{\begin{array}{l}\label{SZC2}
a_x=\lambda pc-\lambda qb,\\
b_x=\lambda qa-\lambda^2 c-\alpha(p^2+q^2)c,\\
c_x=\alpha(p^2+q^2)b+\lambda^2 b-\lambda pa.\end{array}\right.
\end{equation}
By letting
\begin{align}\label{CE2}
a=\sum\limits_{i\ge 0}a_i\lambda^{-2i},\,\, b=\sum\limits_{i\ge 0}b_i\lambda^{-2i-1},\,\, c=\sum\limits_{i\ge 0}c_i\lambda^{-2i-1},
\end{align}
and taking the initial values
\begin{align}
a_0=1,\ b_0=p,\ c_0=q,
\end{align}
we have the following recursion relations from the system \eqref{SZC2}:
%
\begin{equation}
\left\{\begin{array}{l}
a_{i+1,x}=\alpha(p^2+q^2)(qb_i-pc_i)-qc_{i,x}-pb_{i,x},\\
b_{i+1}=pa_{i+1}-\alpha(p^2+q^2)b_i+c_{i,x},\\
c_{i+1}=qa_{i+1}-\alpha(p^2+q^2)c_i-b_{i,x},
\end{array}\right.\  i\ge 0,
\end{equation}
which yields
\begin{align}
\begin{pmatrix}
b_{i+1}\\
c_{i+1}
\end{pmatrix}=L\begin{pmatrix}b_i\\c_i\end{pmatrix},\ L=\begin{pmatrix}L_{11}&L_{12}\\L_{21}&L_{22}\end{pmatrix},\ i\ge 0,
\end{align}
with $L$ being given by
\begin{equation}\left\{\begin{array}{l}\label{ROL}
L_{11}=-\alpha(p^2+q^2)+\alpha p\partial^{-1}q(p^2+q^2)-p\partial^{-1}p\partial,\\
L_{12}=\partial-\alpha p\partial^{-1}p(p^2+q^2)-p\partial^{-1}q\partial,\\
L_{21}=-\partial+\alpha q\partial^{-1}q(p^2+q^2)-q\partial^{-1}p\partial,\\
L_{22}=-\alpha (p^2+q^2)-\alpha q\partial^{-1}p(p^2+q^2)-q\partial^{-1}q\partial.\end{array}\right.
\end{equation}
By imposing the same conditions on the constants of integration as in \eqref{condsofConstantsofIntegration},
 we can determine the sequence $\{a_i,\, b_i,\, c_i|\, i\ge 1\}$ uniquely.
The first two sets can be computed as
\begin{align*}
a_1&=-\frac{1}{2}(p^2+q^2),\
b_1=q_x-\left(\alpha+\frac{1}{2}\right)p(p^2+q^2),\
c_1=-p_x-\left(\alpha+\frac{1}{2}\right)q(p^2+q^2);\\
a_2&=qp_x-pq_x+\left(\alpha+\frac{3}{8}\right)(p^2+q^2)^2,\\
b_2&=-p_{xx}-2\alpha pqp_x-\left(2\alpha+\frac{3}{2}\right)p^2q_x-\left(4\alpha+\frac{3}{2}\right)q^2q_x+\left(\alpha^2+\frac{3}{2}\alpha+\frac{3}{8}\right)p(p^2+q^2)^2,\\
c_2&=-q_{xx}+2\alpha qpq_x+\left(2\alpha+\frac{3}{2}\right)q^2p_x+\left(4\alpha+\frac{3}{2}\right)p^2p_x+\left(\alpha^2+\frac{3}{2}\alpha+\frac{3}{8}\right)q(p^2+q^2)^2.
\end{align*}

Next, let us introduce the auxiliary matrix spectral problems
\begin{align}\label{AP2}
\phi_{t_m}=V^{[m]}\phi,\ V^{[m]}=\lambda (\lambda^{2m+1}W)_++\Delta_m=(\lambda^{2m+2}W)_++(f_m-a_{m+1})e_1,\  m\ge 0,
\end{align}
where $f_m$ satisfy
\begin{align}\label{AT2}
f_{m,x}=2\alpha(pp_{t_m}+qq_{t_m}),\ m\ge 0.
\end{align}
Now the compatibility conditions of \eqref{KNS} and \eqref{AP2}, i.e., the zero curvature equations
\begin{align}
U_{t_m}-V^{[m]}_x+[U,V^{[m]}]=0,\ m\ge 0,
\end{align}
lead to
\begin{equation}\left\{\begin{array}{l}
p_{t_m}=b_{m,x}+\beta(p^2+q^2)c_m-qf_m
=qa_{m+1}-c_{m+1}-qf_m,\\
q_{t_m}=c_{m,x}-\beta(p^2+q^2)b_m+pf_m
=b_{m+1}-pa_{m+1}+pf_m
\end{array}\right. \ m\ge 0.\label{ResultingEqns2}
\end{equation}
Therefore, based on \eqref{ResultingEqns2}, we can solve \eqref{AT2} to get
\begin{align}
f_m=-2\beta a_{m+1},\ m\ge 0.\label{defoff_mof2ndgsh}
\end{align}
Upon plugging \eqref{defoff_mof2ndgsh} into \eqref{ResultingEqns2},
we finally arrive at the hierarchy of generalized Ma equations of KN type:
\begin{align}\label{SH2}
u_{t_m}
=K_m=R\begin{pmatrix}b_{m+1}\\c_{m+1}\end{pmatrix},\
R=\begin{pmatrix}
-(1+2\beta)q\partial^{-1}q&-1+(1+2\beta)q\partial^{-1}p\\
1+(1+2\beta)p\partial^{-1}q&-(1+2\beta)p\partial^{-1}p
\end{pmatrix},\  m\ge 0.
\end{align}
The first nontrivial nonlinear system is given by
\begin{align}
u_{t_1}=\begin{pmatrix}p\\ q\end{pmatrix}_{t_1}=
\begin{pmatrix}
q_{xx}-\frac{1}{2}q^2p_x-\left(\frac{3}{2}+4\beta\right)p^2p_x-(1+4\beta)pqq_x+\left(\frac{1}{4}+\beta\right)\beta q(p^2+q^2)^2\\
-p_{xx}-\frac{1}{2}p^2q_x-\left(\frac{3}{2}+4\beta\right)q^2q_x-(1+4\beta)pqp_x-\left(\frac{1}{4}+\beta\right)\beta p(p^2+q^2)^2.
\end{pmatrix}
\end{align}

\subsection{Bi-Hamiltonian structure and Liouville integrability of the second generalized soliton hierarchy}

In order to establish bi-Hamiltonian structures for the second generalized soliton hierarchy \eqref{SH2},
we shall use the trace identity \eqref{TIF}.
It is direct to calculate that
\begin{equation}\left\{\begin{array}{l}
\mbox{tr}\left(W\frac{\partial U}{\partial \lambda}\right)=-4\lambda a-2pb-2qc,\\
\mbox{tr}\left(W\frac{\partial U}{\partial p}\right)=-4\beta pa-2\lambda  b,\\
\mbox{tr}\left(W\frac{\partial U}{\partial q}\right)=-4\beta qa-2\lambda c.\end{array}\right.
\end{equation}
Therefore, we have
\begin{equation}
\frac{\delta}{\delta u}\int(2\lambda a+pb+qc)dx=\left(\lambda^{-\gamma}\frac{\partial}{\partial \lambda}\lambda^{\gamma}\right)\begin{pmatrix}\lambda b+2\beta pa\\\lambda c+2\beta qa\end{pmatrix}.
\end{equation}
By substituting \eqref{CE2} into the above equation and comparing the coefficients of each power of $\lambda$, we get
\begin{align}
&\frac{\delta}{\delta u}\int(2a_{m+1}+pb_m+qc_m)dx=(\gamma-2m)\begin{pmatrix}b_m+2\beta pa_m\\c_m+2\beta qa_m\end{pmatrix},\,\ m\ge 0
\end{align}
The identity corresponding to $m=1$ tells $\gamma=0$. Thus, we have
\begin{align}
\frac{\delta \mathcal{H}_{m}}{\delta u}
=\begin{pmatrix}b_{m}+2\beta pa_{m}\\c_{m}+2\beta qa_{m}\end{pmatrix},\ m\ge 0,
\end{align}
with
\begin{align}
\mathcal{H}_0=\frac{1}{2}(1+2\beta)\int (p^2+q^2)dx,\, \mathcal{H}_{m}=\int\left(- \frac{2a_{m+1}+pb_{m}+qc_{m}}{2m}\right) dx,\,m\ge 0.
\end{align}

Moreover, noticing that
\begin{equation}
\begin{pmatrix}
b_{m+1}\\c_{m+1}
\end{pmatrix}=N\begin{pmatrix}b_{m+1}+2\beta pa_{m+1}\\
c_{m+1}+2\beta qa_{m+1}\end{pmatrix}
,\ N
=\begin{pmatrix}
1+2\beta p\partial^{-1}q&-2\beta p\partial^{-1}p\\
2\beta q\partial^{-1}q&1-2\beta q\partial^{-1}p
\end{pmatrix}
,\ m\ge 0,
\end{equation}
%
we find that
\begin{align}
u_{t_m}=K_m=J\frac{\delta \mathcal{H}_{m+1}}{\delta u},\ m\ge 0,
\end{align}
with $J=RN$ being given by
\begin{align}
J=\begin{pmatrix}-(1+4\beta)q\partial^{-1}q&-1+(1+4\beta)q\partial^{-1}p\\
1+(1+4\beta)p\partial^{-1}q&-(1+4\beta)p\partial^{-1}p
\end{pmatrix}.
\end{align}

It is obvious that
\begin{equation}
\frac{\delta \mathcal{H}_{m+1}}{\delta u}=\Psi\frac{\delta \mathcal{H}_m}{\delta u},\ \Psi=N^{-1}LN,
\end{equation}
where the inverse operator $N^{-1}$ of $N$ is given by
\begin{equation}
N^{-1}
=\begin{pmatrix}
1-2\beta p\partial^{-1}q&2\beta p\partial^{-1}p\\
-2\beta q\partial^{-1}q&1+2\beta q\partial^{-1}p
\end{pmatrix}.
\end{equation}
Then from $K_{m+1}=\Phi K_m,\ m\ge 0$, and $J\Psi=\Phi J$, we obtain a common recursion operator for the generalized soliton
hierarchy \eqref{SH2}:
\begin{align}
\Phi=\Psi^{\dagger}=N^{\dagger}L^{\dagger}(N^{-1})^{\dagger}
\end{align}
where $N^\dagger$ denotes the adjoint operator of $N$. The operator $\Phi$ can be expressed explicitly as follows:
%
\begin{align}
\Phi=\begin{pmatrix}\Phi_{11}&\Phi_{12}\\ \Phi_{21}&\Phi_{22}\end{pmatrix}\label{recursionoperatorof2ndgsh}
\end{align}
with
\begin{equation}\left\{\begin{array}{l}
\Phi_{11}=2\beta q\partial^{-1}q\partial+2\beta^2 q\partial^{-1} p(p^2+q^2)+2\beta(1+2\beta)q\partial^{-1}q\partial q\partial^{-1}p\\
\qquad\quad-\beta (p^2+q^2)[1+(1+2\beta)q\partial^{-1}p]-(1+2\beta)(1-2\beta q\partial^{-1}p)\partial p\partial^{-1}p,\\
\Phi_{12}=(1-2\beta q\partial^{-1}p)[\partial-(1+2\beta)\partial p\partial^{-1}q]\\
\qquad\qquad-\beta(1+2\beta)[(p^2+q^2)-2q\partial^{-1}q\partial]q\partial^{-1}q+2\beta^2q\partial^{-1}q(p^2+q^2),\\
\Phi_{21}=-(1+2\beta p\partial^{-1}q)[\partial+(1+2\beta)\partial q\partial^{-1}p]\\
\qquad\quad+\beta(1+2\beta)[(p^2+q^2)-2p\partial^{-1}p\partial]p\partial^{-1}p-2\beta^2 p\partial^{-1}p(p^2+q^2),\\
\Phi_{22}=2\beta p\partial^{-1}p\partial-2\beta^2p\partial^{-1}q(p^2+q^2)-2\beta(1+2\beta)p\partial^{-1}p\partial p\partial^{-1}q\\
\qquad\quad-\beta(p^2+q^2)[1-(1+2\beta)p\partial^{-1}q]-(1+2\beta)(1+2\beta p\partial^{-1}q)\partial q\partial^{-1}q.\end{array}\right.
\end{equation}

Obviously, we can similarly show that all members of the generalized soliton hierarchy \eqref{SH2} are bi-Hamiltonian:
\begin{equation}
u_{t_m}=K_m=J\frac{\delta \mathcal{H}_{m+1}}{\delta u}=M\frac{\delta \mathcal{H}_{m}}{\delta u},\ m\ge 0,\label{biHamiltonainStructuresof2ndgsh}
\end{equation}
where the second Hamiltonian operator $M$ is given by 
%
\begin{equation}
M=\Phi J=\begin{pmatrix}M_{11}&M_{12}\\ M_{21}&M_{22}\end{pmatrix}
\end{equation}
with the entries of $M$ being defined by
\begin{equation}\left\{\begin{array}{l}
M_{11}=\partial-2\beta q\partial^{-1}p\partial+2\beta(1-2\beta q\partial^{-1}p)\partial p\partial^{-1}q\\
\qquad\quad+2\beta^2 q \partial^{-1}q(p^2+q^2)+2\beta^2[(p^2+q^2)-2q\partial^{-1}q\partial]q\partial^{-1}q,\\
M_{12}=\beta(1-2\beta q\partial^{-1}p)[(p^2+q^2)-2\partial p\partial^{-1}p]\\
\qquad\quad-2\beta^2[(p^2+q^2)-2q\partial^{-1}q\partial]q\partial^{-1}p-2\beta q\partial^{-1}q\partial,\\
M_{21}=\beta(1+2\beta p\partial^{-1}q)[2\partial q\partial^{-1}q-(p^2+q^2)]\\\qquad\quad+2\beta^2[2p\partial^{-1}p-(p^2+q^2)]p\partial^{-1}q+2\beta p\partial^{-1}p\partial,\\
M_{22}=\partial+2\beta p\partial^{-1}q\partial-2\beta(1+2\beta p\partial^{-1}q)\partial q\partial^{-1}p\\
\qquad\quad+2\beta^2p\partial^{-1}p(p^2+q^2)+2\beta^2[(p^2+q^2)-2p\partial^{-1}p\partial]p\partial^{-1}p.\end{array}\right.
\end{equation}
These bi-Hamiltonian structures show the Liouville integrability of the generalized Ma soliton hierarchy of KN type \eqref{SH2}.
Further, it follows that there are infinitely many commuting common symmetries and conserved functionals
\begin{align}
[K_l,K_m]=K_l'(u)[K_m]-K_m'(u)[K_l]=0,\ l, m\ge 0
\end{align}
and
\begin{equation}\left\{\begin{array}{l}
\{\mathcal{H}_l,\mathcal{H}_m\}_{J}=\int ( \frac{\delta\mathcal{H}_l}{\delta u})^T J\frac{\delta \mathcal{H}_m}{\delta u}dx=0,\\
\{\mathcal{H}_l,\mathcal{H}_m\}_{M}=\int ( \frac{\delta\mathcal{H}_l}{\delta u})^T M\frac{\delta \mathcal{H}_m}{\delta u}dx=0,\end{array}\right. \  l, m\ge 0.
\end{equation}
\begin{rem}
By taking the ansatz $\beta=0$ for the generalized matrix spectral problem \eqref{KNS},
 the
recursion operator \eqref{recursionoperatorof2ndgsh} and the bi-Hamiltonian structures \eqref{biHamiltonainStructuresof2ndgsh}
of the corresponding generalized soliton hierarchy \eqref{SH2}
 are reduced to the ones
 of the Ma soliton hierarchy of KN type
 presented in \cite{MA2}. 
\end{rem}

\section{Conclusion and discussion}
In this paper, we proposed two new matrix spectral problems associated with so(3,$\mathbb{R}$), that is, the generalized AKNS spectral problem and the generalized KN spectral problem. We derived two hierarchies of soliton equations by the standard procedure
using the zero curvature formulation,
together with their recursion operators and bi-Hamiltonian structures, and finally proved their integrability in the sense of Liouville.
The discussed two generalized matrix spectral problems and their corresponding results are reduced to the ones presented in \cite{MA1} and \cite{MA2}, upon making the choice $\beta=0$.

In \cite{MA3}, Ma et al. successfully generalized the classical WKI matrix spectral problem associated with sl(2,$\mathbb{R}$):
\begin{align}\label{WKI}
\phi_x=\lambda(e_1'+pe_2'+qe_3')\phi
\end{align}
to the following matrix spectral problem associated with so(3,$\mathbb{R}$):
\begin{align}\label{GWKI}
\phi_x=\lambda(e_1+pe_2+qe_3)\phi,
\end{align}
by taking the same linear combination of the basis matrices.
However, it remains a problem how to generalize the matrix spectral problem \eqref{GWKI}
 associated with so(3,$\mathbb{R}$).
We hope there will be
some generalized
matrix spectral problem for \eqref{GWKI}, which can yield a generalized soliton hierarchy as we did above.


\section*{Acknowledgement}
This work was supported by the National Natural Science Foundation of China under the grants 11271266, 11271008,
11371323, 11371326 and 61072147, NSF under the grant DMS-1301675, Natural Science Foundation of Shanghai (Grant No. 11ZR1414100), Zhejiang Innovation Project of China (Grant No. T200905), and the First-class Discipline of Universities in Shanghai and Shanghai Univ. Leading Academic Discipline Project (No. A.13-0101-12-004). The authors would also like to thank S. Manukure and W.Y. Zhang for their valuable discussions in DE seminars at University of South Florida.

\end{document}